\documentclass{aa}  

\usepackage{graphicx}
\usepackage{caption}
\usepackage{natbib}
\bibpunct{(}{)}{;}{a}{}{,} 
\usepackage{threeparttable}
\usepackage{float}
\usepackage{xcolor}
\definecolor{darkgreen}{RGB}{0,125,0}

\usepackage{txfonts}
\usepackage[colorlinks=true, linkcolor=blue, citecolor=blue]{hyperref}
\expandafter\def\expandafter\normalsize\expandafter{%
    \normalsize%
    \setlength\abovedisplayskip{4pt}%
    \setlength\belowdisplayskip{8pt}%
    \setlength\abovedisplayshortskip{-8pt}%
    \setlength\belowdisplayshortskip{2pt}%
}

\begin{document}

   \title{Discovery of the first outbursting hot subdwarf binary: ZTF J0007+4804}

    \author{E. Stringer
    \inst{1}
    \and
    T. Kupfer\inst{1,2}
    \and
    K. Deshmukh\inst{3}
    \and
    T. Maccarone\inst{2}
    \and
    I. Jackson\inst{4}
    \and
    A. Kosakowski\inst{2,5}
    \and
    C. W. Bradshaw\inst{1}
    \and \\
    A. Brown\inst{1}
    \and
    M. Dorsch\inst{6}
    \and
    A. Picco\inst{3}
    \and
    V.S. Dhillon\inst{7,8}
    \and
    S. Poshyachinda\inst{9}
    \and
    S. Awiphan\inst{9}
}

   \institute{Hamburger Sternwarte, University of Hamburg, Gojenbergsweg 112, 21029 Hamburg, Germany\\
        \email{eric.stringer@uni-hamburg.de}
        \and
        Department of Physics \& Astronomy, Texas Tech University, Box 41051, Lubbock, TX, 79409-1051, USA
        \and
        Institute of Astronomy, KU Leuven, Celestijnenlaan 200D, B-3001 Leuven, Belgium
        \and
        Astronomy and Physics Department, Lycoming College, Williamsport, PA 17701, USA
        \and
        Department of Physics and Astronomy, University of North Carolina at Chapel Hill, Chapel Hill, NC 27599, USA
        \and
        Institut f\"ur Physik und Astronomie, Universit\"at Potsdam, D-14476 Potsdam-Golm, Germany
        \and
        Astrophysics Research Cluster, School of Mathematical and Physical Sciences, University of Sheffield, Sheffield, S3 7RH, UK
        \and
        Instituto de Astrof´ısica de Canarias, E-38205 La Laguna, Tenerife, Spain
        \and
        National Astronomical Research Institute of Thailand (Public Organization), 260 Moo 4, Donkaew, Mae Rim, Chiang Mai 50180, Thailand
}

   \date{submitted April 13, 2026}


\abstract
    {Hot subdwarf binaries with white dwarf companions with orbital periods of less than two hours are progenitor candidates for massive single white dwarfs as well as a variety of thermonuclear explosions.}
    {Our aim is to determine the binary properties of the hot subdwarf -- white dwarf system ZTF J000742.62+480414.51, model its future evolution, and characterize the brightening events seen in TESS photometry.}
    {Using data from ZTF and TESS, we performed a Lomb Scargle analysis to find the orbital period and the period of the brightening events. Analysis of time-resolved spectroscopy was combined with light curve modeling to determine the effective temperature, surface gravity, and radius of the primary star, the masses of both stars, and to confirm the presence of an accretion disk. X-ray observations were performed with Swift, and MESA modeling was used to find the future evolution of the system. The kinematics of the system were also calculated.}
    {ZTF J000742.62+480414.51 consists of an accreting $0.48\pm0.01\,M_\odot$ white dwarf with a $0.42\pm0.01\,M_\odot$ B-type hot subdwarf acting as a donor. The system exhibits SU UMa type dwarf nova outbursts with a recurrence time of $P_{\mathrm{out}} \approx 9$ days. No X-rays were detected, with an upper limit on the X-ray luminosity of about $3\times10^{31}$ erg/sec. The system lies in the Galactic thin disk, and has an orbital period of $P_{\mathrm{orb}} = 108.72\pm0.01$ minutes. The system has likely formed from a main sequence binary with component masses $\gtrsim2\,\mathrm{M_{\odot}}$ and will likely merge into a single white dwarf, but a thermonuclear explosion cannot be ruled out.}
    {ZTF J000742.62+480414.51 consists of a low mass white dwarf actively accreting hydrogen rich material from a B-type hot subdwarf, and is the first hot subdwarf -- white dwarf system discovered that produces dwarf nova outbursts.}




   \maketitle
%
\section{Introduction}
\noindent
Hot subdwarf B stars (sdBs) are subluminous stars with temperatures between 20\,000 K and 40\,000 K that reside on the extreme horizontal branch of the Hertzsprung-Russell (HR) diagram. These stars are thought to be the helium core burning remnants of red giants which have lost most of their hydrogen envelope, leaving only a thin hydrogen layer surrounding the core \citep{Heber_1986,Heber_2009,Heber_2016}. Between one half \citep{Napiwotzki_2004,Copperwheat_2011} and two thirds \citep{Maxted_2001} of all sdBs are in binaries with periods of less than ten days, leading to the conclusion that the loss of the hydrogen envelope is often due to interaction with a binary companion. Recent studies even propose that binary evolution is required to form hot subdwarfs \citep{pelisoli2020}. To produce binaries with such short orbital periods, these systems must have experienced at least one common envelope (CE) phase (where the companion orbits inside the atmosphere of the proto-sdB) near the time that the sdB began burning helium in its core \citep{Han_2002,Han_2003}.

After the CE phase, the stars' orbit will continue to shrink due to angular momentum loss via gravitational wave emission. In some cases, the sdB can overflow its Roche lobe, donating material to its companion via accretion. For sdBs with white dwarf (WD) companions, transfer of the sdB's hydrogen envelope can begin at orbital periods of 50 -- 150 minutes, with transfer of the helium core starting at $P_{\mathrm{orb}} \lesssim$ 40 minutes. For mass transfer to begin, the sdB must reach these orbital periods before it turns into a WD. This puts an upper limit on the range of post-CE orbital periods that will allow for future envelope and core mass transfer. For typical compact sdB-WD systems ($M_{\mathrm{sdB}} = 0.32 - 0.55\,\mathrm{M_\odot}$, $M_{\mathrm{WD}} = 0.75\,\mathrm{M_\odot}$), the upper limits are expected to be $\approx$ 185 minutes and $\approx$ 165 minutes for mass transfer of the envelope and the core respectively. For details on sdB-WD post CE evolution, see \cite{Bauer_2021}.

For these reasons, hot subdwarf-WD binaries with short orbital periods are potential Type Ia supernova progenitors through the double detonation channel. In this channel, the WD accretes helium-rich material from a companion star, building up a helium layer on its surface. After accreting $\approx$ 0.1 $\mathrm{M_{\odot}}$ the degenerate portion of the helium layer will ignite, compressing the WD and causing a carbon fusion chain reaction in the core which in turn detonates the star \citep{Livne_1990,Livne_1995,Fink_2010,Woosley_2011,Wang_2012,Shen_2014,Wang_2018}. So far, there are two hot subdwarf--WD systems that are thought to be double detonation progenitors: CD--$30^{\circ}$11223 and PTF1 J223857.11+743015.1 \citep{Vennes_2012,Geier_2013,Kupfer_2022}, though neither of these are currently in their mass transfer phase. It is still unclear how double detonations contribute to the overall population of Type Ia supernovae, so understanding the nature of their progenitors and similar non-progenitor systems may give us insight into this puzzle.

There are currently three known hot subdwarf--WD systems undergoing stable Roche lobe overflow: ZTF J213056.71+442046.5, ZTF J205515.98+465106.5, and SMSS J192054.50--200135.5 \citep{Kupfer_2020a,Kupfer_2020b,Li_2022}. All three were confirmed as mass transferring systems from the detection of eclipses of the hot subdwarf by the accretion disk, although no direct accretion signatures have been found so far \citep[e.g.][]{2019ATel12847....1R,2020ATel13444....1R,2022Mereghetti,2023Deshmukh}. None of the hot subdwarfs in these systems are suspected of being helium core burning objects. ZTF J213056.71+442046.5 and J205515.98+465106.5 are in a short hydrogen shell burning phase near the end of their hot subdwarf stage, while SMSS J192054.50--200135.5 is suspected of being a stripped AGB star also undergoing shell burning. None of these objects are thought to have formed from low mass main sequence stars like most other hot subdwarfs, but rather from progenitors with masses $\gtrsim2\,\mathrm{M_{\odot}}$ which would make these systems younger than the general hot subdwarf population with canonical hot subwarf masses \citep{Han_2003}.

In this paper we present a full analysis of ZTF J000742.62+480414.51 (hereafter ZTF J0007+4804), a new mass transferring sdB-WD binary system. This system was first reported by \cite{Geier_2019} as part of an all-sky catalogue of $\sim$ 40\,000 hot subdwarf candidates in Gaia DR2, and later identified as an ellipsoidal hot subdwarf--WD binary candidate with a 1.8 hr orbital period by \cite{Barlow_2022}, using data from TESS. In section \ref{sec:data} we describe the photometric and spectroscopic data used for this paper and in section \ref{sec:methods} we describe the analysis, our approach to modeling the light curve, and the analysis of the X-ray data from Swift. Section \ref{sec:results} lists the results of the study. In section \ref{sec:discussion} we discuss the unique properties of the system, how this system compares to other Roche-lobe filling sdB-WD binaries, as well as its kinematics and evolutionary history and future. We conclude and summarize in section \ref{Conclusion}.


      \begin{figure}
   \centering
   \includegraphics[width=\hsize]{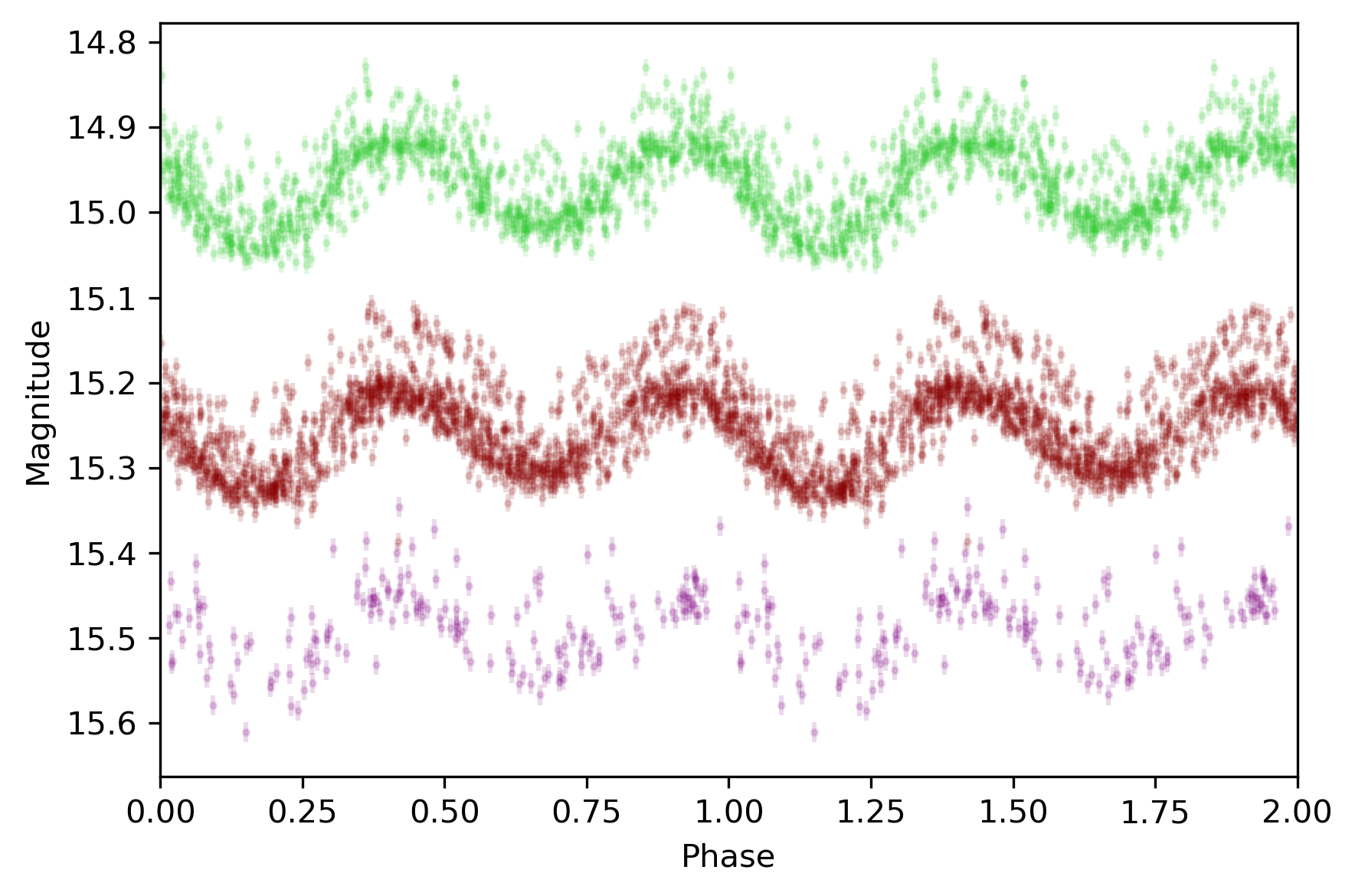}
      \caption{ ZTF light curve of ZTF J0007+4804 folded over the orbital period of 108.72 minutes for (from top to bottom) ZTF $g$, ZTF $r$, and ZTF $i$ bands. The data are repeated for clarity.
              }
         \label{Phase_Folded_Lightcurve}
   \end{figure}

      \begin{figure}
   \centering
   \includegraphics[width=\hsize]{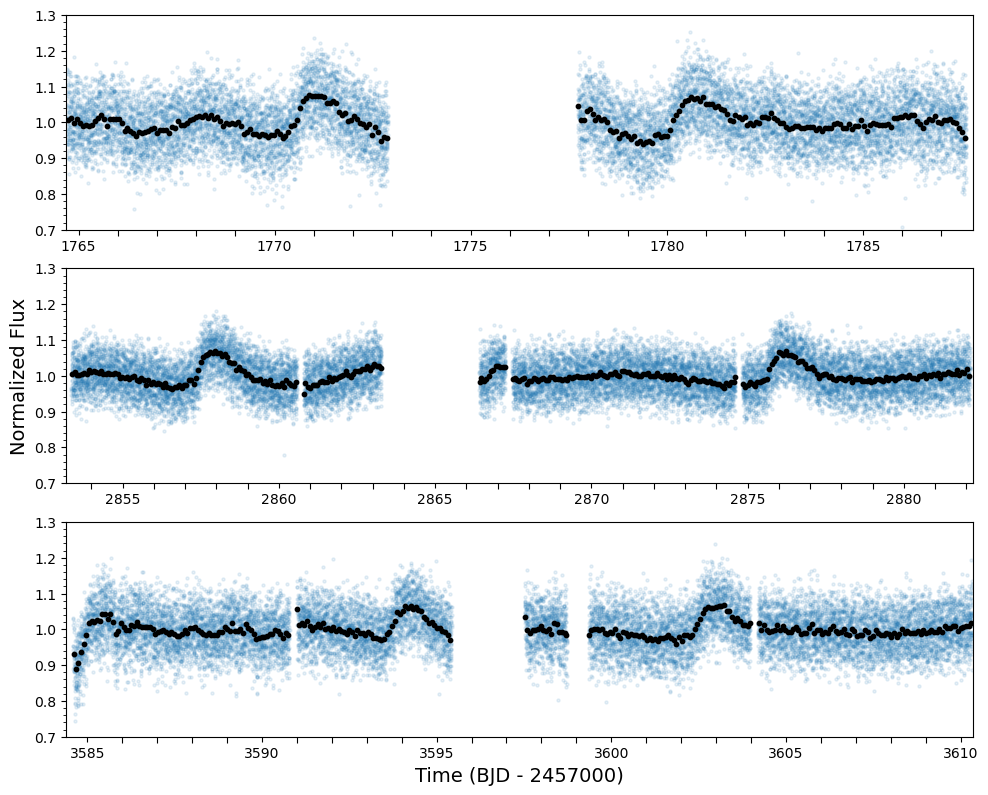}
      \caption{TESS light curve of ZTF J0007+4804. The data were taken in three different epochs, which are shown above. The black dots are the data points binned over a single orbital period.
              }
         \label{Tess_Lightcurve}
   \end{figure}

\section{Data}\label{sec:data}

\noindent
Photometry for the analysis of ZTF J0007+4804 was taken by several telescopes. The system was observed as part of the Zwicky Transient Facility (ZTF) public survey \citep{Graham_2019,Bellm_2019} in the $g$, $r$, and $i$ bands with 30 second exposures, which were taken between May 2018 and February 2024 for a total of 2249 data points. Image processing of ZTF data is described in full detail in \cite{Masci_2019}. 

We also used 120 second cadence data from the Transiting Exoplanet Survey Satellite (TESS; \citealt{Ricker_2015}). TESS observed ZTF J0007+4804 (TIC 201736330) in sectors 17, 57, and 84 between October 2019 and October 2024 for a total of 47\,104 data points. The reduced data were retrieved from the Barbara A. Mikulski Archive for Space Telescopes (MAST)\footnote{\url{https://mast.stsci.edu}}.


We obtained 109 minutes of high-speed SDSS r-band data with 5-second cadence using the Argos instrument on the 2.1-meter Otto Struve telescope at the McDonald observatory on 2021 September 04. The McDonald data was reduced using standard IRAF procedures for image bias, dark current, and flat-field corrections with image calibration data obtained on the same night as our observations. We extracted the light curve of ZTF J0007+4804 and four nearby field stars using the \textsc{phot} task within the \textsc{daophot} package of IRAF \citep{tody1986,tody1993}. We applied a variable circular aperture radius for flux extraction based on the average PSF FWHM in each individual image. We then calibrated the light curve of ZTF J0007+4804 using the weighted-mean combined light curve created from the light curves of four nearby non-variable field stars.

Finally, we observed a single orbit in the $u$, $g$, and $i$ bands with 1.38 second exposures with the Thai National Observatory's Thai National Telescope, which were taken on the night of December 10th, 2024. The high-speed imaging photometer ULTRASPEC was used \citep{Dhillon_2014}, and data reduction was done using the HiPERCAM pipeline \citep{Dhillon_2021}.


ZTF J0007+4804 was observed in X-rays by the Swift X-ray Telescope (XRT) on March 19th, 2025. Two separate exposures were taken with the first being 495 seconds, and the second lasting 1395 seconds for a total of 1890 seconds of exposure time. Further information on the Swift XRT data is given by \cite{Burrows_2000} and \cite{Hill_2000}.

Time series spectroscopic observations were taken with the Keck observatory's Echellette Spectrograph and Imager (ESI) \citep{Sheinis_2002}, as well as with the Lick Observatory's Shane Kast blue spectrograph. The ESI spectra were reduced using the \texttt{MAKEE}\footnote{\url{https://sites.astro.caltech.edu/~tb/makee/}} pipeline following the standard procedure: bias subtraction, flat fielding, sky subtraction, order extraction, and wavelength calibration. The Shane Kast blue spectra were reduced using PypeIt \citep{Prochaska_2020}. Overall, 29 spectra covering a single orbit taken on December 7th, 2020 were used from ESI, and 69 spectra from three separate nights (July 30th, 2019, September 5th, 2019, and September 6th, 2019) were used from Shane Kast blue. 



  \begin{figure}
   \centering
   \includegraphics[width=\hsize]{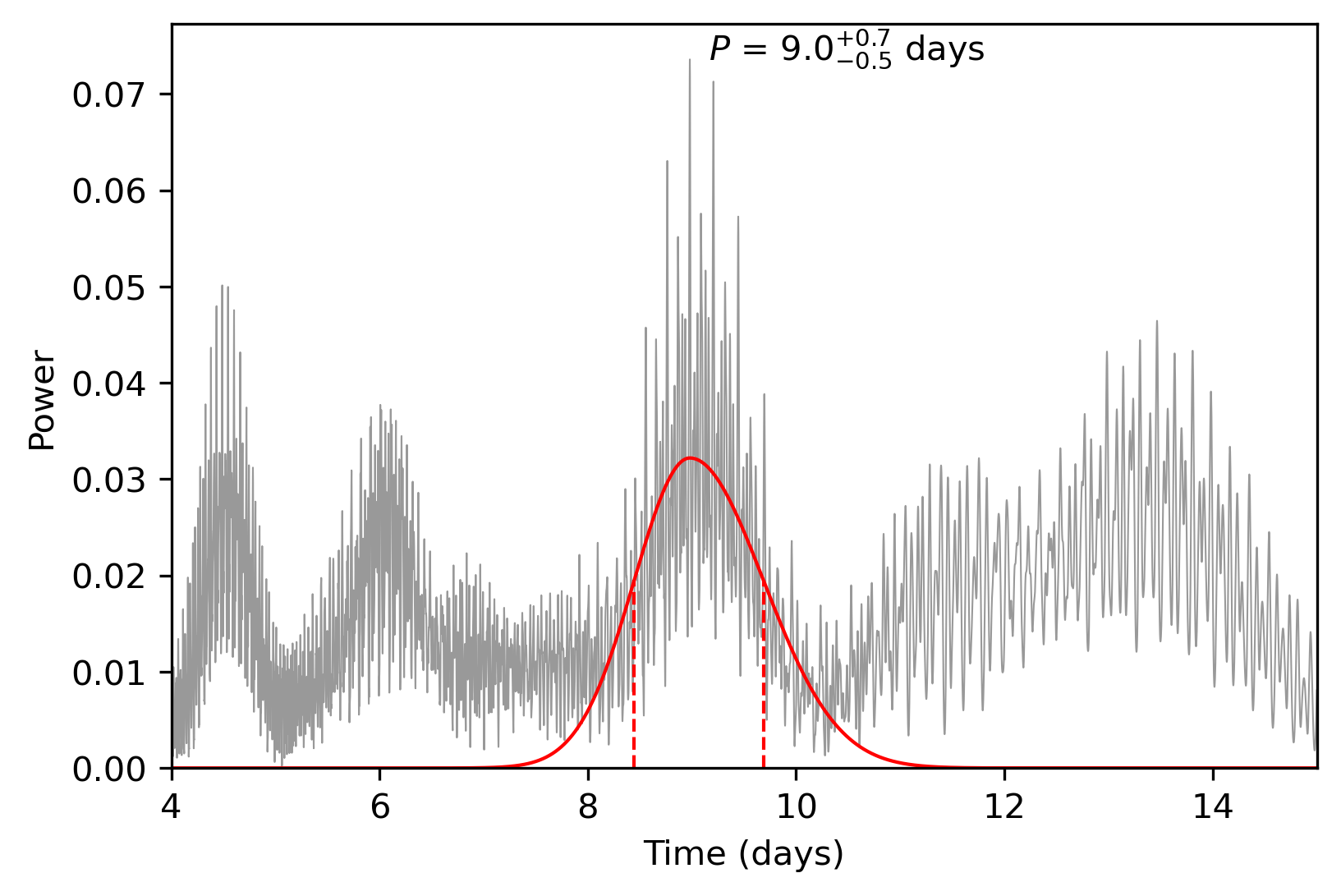}
      \caption{Lomb-Scargle periodogram of the TESS light curve. The red line is the split Gaussian fit of the peak profile using the maximum peak as the mean. The dotted red lines show the 1$\sigma$ uncertainties.}
         \label{Tess_Lomb_Scargle}
   \end{figure}

\section{Methods}\label{sec:methods}

\subsection{Photometric and Spectroscopic Analysis}
\noindent

To determine the presence of the binary signal in the ZTF photometry, we computed a Lomb-Scargle periodogram \citep{Lomb_1976, Scargle_1982} on each of the three available ZTF $g$, $r$, and $i$ bands individually, using the Astropy Lomb-Scargle periodogram module \citep{Astropy_2013,Astropy_2018}. Since the light curve variability is expected to arise from ellipsoidal modulation, which results in two maxima and two minima per orbit, we used two sinusoidal terms (nterms = 2) when computing the power spectrum to derive the orbital period and band-dependent amplitudes.

To further refine the binary period and its uncertainty since the initial reporting of 1.8~hr by \cite{Barlow_2022} using TESS sector 17, we included the latest two TESS sectors (57 and 84) and computed a discrete Fourier transform (DFT) and Monte-Carlo estimation of uncertainties on the combined light curve using Period04 \citep{Lenz_2005}. To analyze the low-frequency regime of the TESS data in order to isolate any longer period signals (see section~\ref{sec:results}), we compute a Lomb-Scargle periodogram in the frequency range corresponding to periods between 4 and 15 days, using three sinusoidal terms (nterms = 3). 

The spectra are single-lined with the sdB outshining the WD. To measure radial velocities of the sdB, we fit a Voigt profile to each individual line of hydrogen and neutral helium and measured the wavelength shift from the rest wavelength in each ESI spectrum. The velocities of each line were calculated using the non-relativistic Doppler equation $v = c\, \frac{\Delta \lambda}{\lambda_0}$, then averaged to get a velocity for each spectrum. Assuming a circular orbit, we then performed a sine-fit of the velocities of the ESI data to find the radial velocity semi-amplitude $k_{\mathrm{sdB}}$ and system velocity $\gamma$.


To obtain the atmospheric parameters of $T_{\mathrm{eff}}$, $\log g$, helium abundance ($\log y = \log\frac{n(\mathrm{He})}{n(\mathrm{H})}$), and projected rotational velocity ($v_{\mathrm{rot}}\sin i$), we coadded the radial velocity-corrected spectra for each individual data set from ESI and Kast and fit a model spectrum with the Interactive Spectral Interpretation System (ISIS) \citep{Houck_2000}. This method is described in detail in \cite{Irrgang_2014_phd}. The model grid is based on \textsc{Atlas12} \citep{kurucz_1996} atmospheres, \textsc{Detail} \citep{Giddings_1981} departure coefficients, and \textsc{Surface} \citep{Giddings_1981} model spectra, and thus uses a hybrid local thermal equilibrium (LTE) / non-LTE approach, the details of which can be found in \citet{przybilla_2011}. The specific models used in this work are described in detail in \cite{Heber_2026} and \cite{Dawson_2026}. The Shane Kast blue spectra were used to get $T_{\mathrm{eff}}$, $\log g$, and $\log y$ and the ESI spectra were used to get $v_{\mathrm{rot}}\sin i$ using the results from Kast.

Finally, $T_{\mathrm{eff}}$, $\log g$, and $\log y$ were combined with Gaia DR3 \citep{Gaia_Collaboration_2023} parallax measurements in a spectral energy distribution (SED) fit to determine the radius of the sdB ($R_{\mathrm{sdB}}$). Parallax zero-point offsets were corrected following \cite{Lindegren_2021}, and the reddening law was taken from \cite{Fitzpatrick_2019}. An SED fits a model spectrum to photometric data points taken from a wide range of bands from several surveys to determine the luminosity of the system. The photometry used for the SED was taken at random phases. This is accounted for by a quadratic addition of a generic uncertainty to all magnitudes $\delta_{\mathrm{excess}}$ such that the final reduced $\chi^2=1$. The angular diameter $\Theta$ and the monochromatic color excess $E(44-55)$ were free parameters. A full treatment of this method is given by \cite{Heber_2018}. 

The photometric data used in the SED fit are sourced from: Two Micron All-Sky Survey (2MASS) \citep{Cutri_2003, Skrutskie_2006}, Galaxy Evolution Explorer (GALEX) \citep{Martin_2005,Bianchi_2017}, Gaia early data release 3 \citep{Gaia_Collaboration_2016,Gaia_Collaboration_2021}, Panoramic Survey Telescope and Rapid Response System (Pan-STARRS) data release 2 \citep{Magnier_2020}, United Kingdom Infrared Telescope Hemisphere Survey (UHS) \citep{Dye_2018,Schneider_2025}, and the Wide-field Infrared Survey Explorer (WISE) unWISE catalog \citep{Schlafly_2019}.


  \begin{figure}
   \centering
   \includegraphics[width=\hsize]{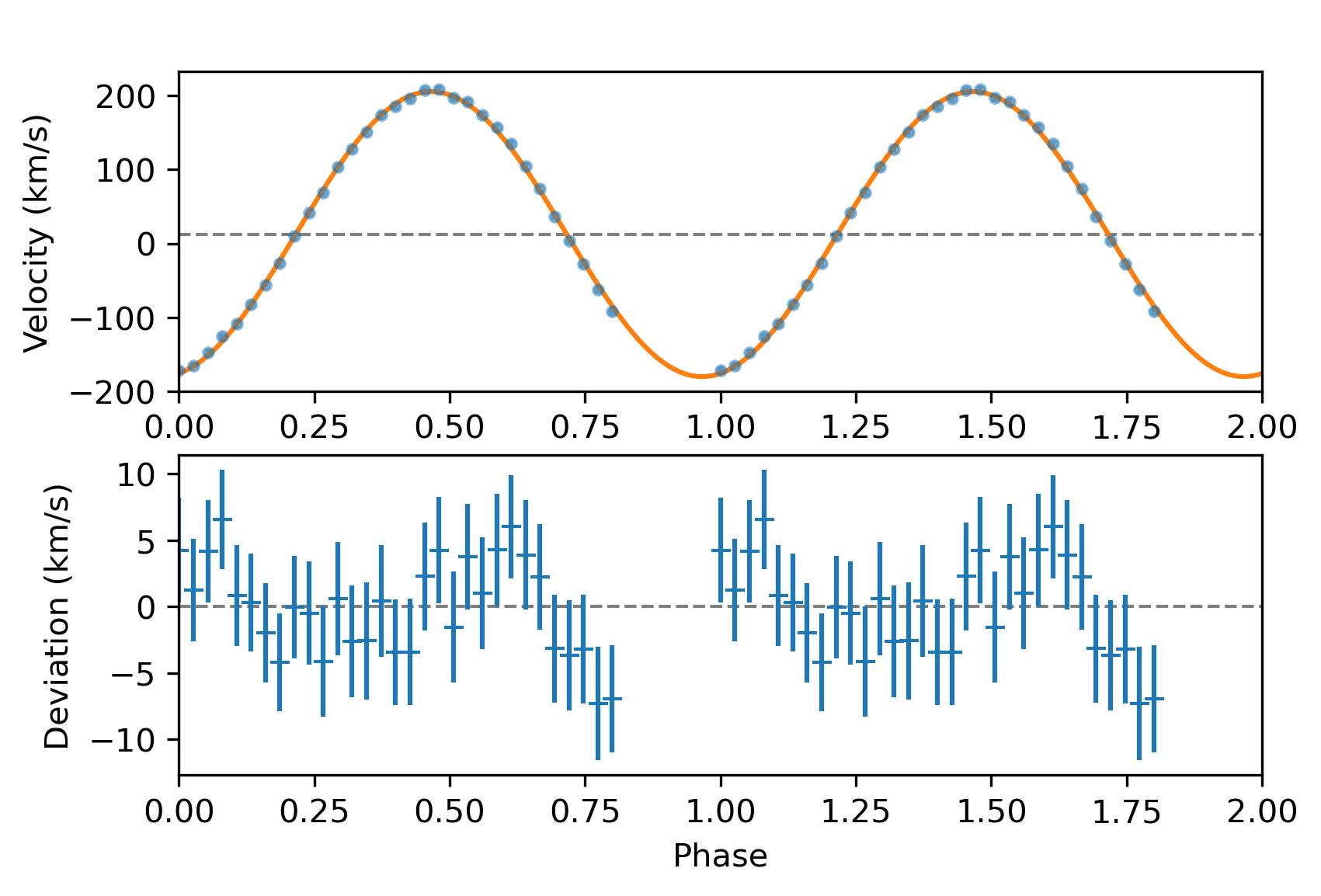}
      \caption{Top panel: (Blue) radial velocity amplitudes plotted against the orbital phase. (Orange) sine fit of the radial velocity amplitudes. (Dashed) the mean of the sine fit, which is taken to be the system velocity $\gamma$. Bottom panel: the residuals of the top panel.}
         \label{Velocities}
   \end{figure}

       \begin{figure*}
   \centering
   \includegraphics[width=\textwidth]{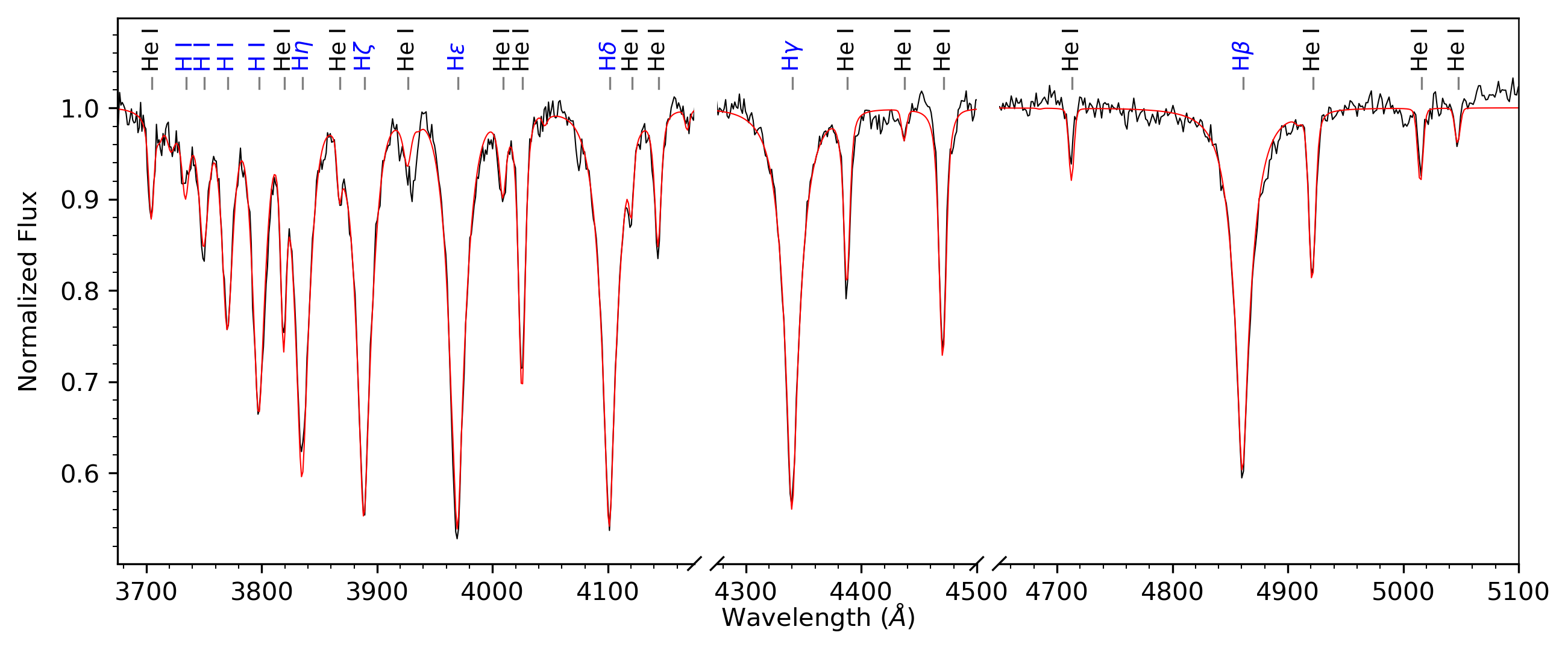}
    \caption{Model spectrum (red) fit against the radial velocity-corrected average spectrum from Shane Kast blue (black). The hydrogen Balmer series is marked in blue, while the neutral helium lines are marked in black.}
   \label{Spectrum}
    \end{figure*}

\subsection{Light Curve Modeling}
\label{Light curve Modeling}
To determine precise system properties we employ the light curve modeling code LCURVE \citep{Copperwheat_2010}. LCURVE creates a grid of points on the surface of each object in the binary and calculates the flux at each point. LCURVE then creates a light curve from this flux using the photometry and the previously calculated orbital period as a reference. Photometry from TESS, McDonald, and the Thai National Telescope was used together to create a single light curve model. Gravity-darkening and limb-darkening coefficients were taken from \cite{Claret_2011} and \cite{Claret_2017} for the SDSS and TESS filters respectively.

We assume a circular orbit, that the system is tidally locked, and that the sdB is Roche-lobe filling. The assumption of a Roche-lobe filling system is driven by the need for an accretion disk to explain periodic brightening events seen in the TESS data (see sections \ref{sec:results}  and \ref{sec:discussion}). For simplicity, we also assume a simple disk model with a constant temperature and a minimum radius equal to $R_{\mathrm{WD}}$. A uniform disk thickness of 0.01 $R_{\odot}$ was also assumed. To find the best-fit model, we used a Markov chain Monte Carlo (MCMC) method implemented in the emcee package \citep{Foreman-Mackey_2013} using $\log g$, $v_{\mathrm{rot}}\sin i$, $R_{\mathrm{sdB}}$, $K_{\mathrm{sdB}}$, and the flux ratio of the disk to the sdB $\frac{F_{\mathrm{disk}}}{F_{\mathrm{sdB}}}$ (see appendix~\ref{A}) as priors, and holding $T_{\mathrm{eff}}$ of the sdB fixed at the value found from the spectral fit. $T_{\mathrm{eff}}$ of the WD was held fixed at an assumed 20\,000 K, and $R_{\mathrm{WD}}$ was constrained by the WD mass-radius relation \citep{Fontaine_2001}.  We then fit $M_{\mathrm{sdB}}$, $M_{\mathrm{WD}}$, the system inclination $i$, the outer disk radius $R_{\mathrm{disk}}$, and the disk temperature $T_{\mathrm{disk}}$. $R_{\mathrm{sdB}}$, $\log g$, $K_{\mathrm{sdB}}$, and the orbital separation $a$ were derived from the fitted parameters using pylcurve \citep{Brown_2022}



       \begin{figure}
   \centering
   \includegraphics[width=\hsize]{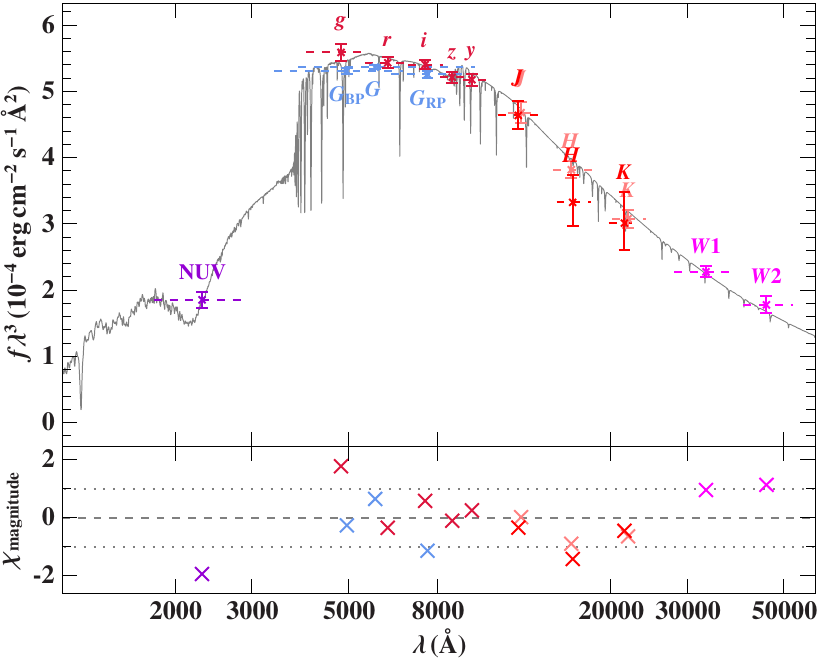}
      \caption{SED fit of ZTF J0007+4804 with photometric data points plotted against the best-fit model spectrum. Data points are color coded by survey, and are sourced from (left to right): GALEX (purple), Gaia EDR3 (blue), Pan-STARRS (crimson), 2MASS (red), UHS (light red), and unWISE (pink). The bottom panel shows the uncertainty-scaled residuals $\chi$ in magnitudes. }
         \label{SED}
   \end{figure}
   

\subsection{X-ray Analysis}
The live time for the Swift XRT exposure is 1873 seconds. The XRT data show 3 photons within a 20 pixel radius extraction region around the source position. A 100 pixel radius background region shows 91 photons, yielding a background within the source region of 3.6 photons, giving a source count rate of -0.6$\pm$1.9 photons. Using Table 3 of \cite{Kraft_1991}, the 99\% confidence level upper limit on the number of counts is 7.3 photons, and the upper limit on the count rate is 3.9$\times10^{-3}$ cts/sec. Taking the foreground absorption from the HEASARC W3NH tool of $9.7\times10^{20}$ H cm$^{-2}$, and a $\Gamma=2$ power law for a spectral shape, this corresponds to an unabsorbed flux of $1.7\times10^{-13}$ erg/sec/cm$^2$ corresponding to an X-ray luminosity of about $4.5\times10^{31}$ erg/sec at a distance of 1.5 kpc.  This value is higher than the quiescent luminosities of typical dwarf novae, and of most dwarf novae at most times of their outbursts \citep{2006csxs.book..421K}.  We also check for power law indices of 1.5 and 2.5 and find that both give flux estimates about 20\% lower, and so the conclusions are robust to reasonable assumptions about spectral indices.


\section{Results}
\label{sec:results}
\noindent
An orbital period of $P = 108.72 \pm 0.01$ minutes was found from the photometric measurements. When phase folding on the peak period, the ZTF light curves show the uneven minima emblematic of ellipsoidal modulation, indicating that the orbital period is the peak signal of $P_{\mathrm{orb}} = 108.72 \pm 0.01$ minutes. A scattering of points biased toward higher flux can also be seen in the phase folded light curve which is not explained by the orbital variation (Figure~\ref{Phase_Folded_Lightcurve}). Looking at the TESS light curve, signals on timescales much longer than the orbital period can be seen with a particularly strong brightening event occurring roughly every nine days (Figure~\ref{Tess_Lightcurve}). The Lomb-Scargle analysis of the TESS data gives a period for these brightening events of $P_{\mathrm{out}} = 9^{+0.7}_{-0.5}$ days. The 1$\sigma$ uncertainties for the brightening period were found by fitting a split Gaussian function to the region around the peak signal, using the location of the peak value as the mean (Figure~\ref{Tess_Lomb_Scargle}). 


From the sine-fit of the radial velocities to the ESI data we find a velocity semi-amplitude of $K_{\mathrm{sdB}} = 193.8 \pm 1.4$ km/s and a system velocity of $\gamma = 12.7 \pm 0.6$ km/s (Figure~\ref{Velocities}). The spectral modeling shows $T_{\mathrm{eff}} = 23\,500 \pm 800$ K, $\log g = 5.26 \pm 0.11$, $v_{\mathrm{rot}}\sin i = 131.86 \pm 1.44$ km/s, and a nearly solar helium abundance of $\log y = -0.99 \pm 0.07$ (Figure~\ref{Spectrum}). The SED fit gives an sdB radius of $R_{\mathrm{sdB}} = 0.296 \pm 0.016$ $\mathrm{R_{\odot}}$ (Figure~\ref{SED}) and a generic excess noise $\delta_{\mathrm{excess}}=0.005\,\mathrm{mag}$. From the calculation of the flux ratio we find $\frac{F_{\mathrm{disk}}}{F_{\mathrm{sdB}}} \approx 3\%$. As we have no disk models in hand, the SED is a single component fit, meaning the 3\% flux contribution of the disk is being attributed to the sdB in the form of a larger radius. To reflect the uncertainty from the additional disk contribution, we increase the uncertainty in the radius to $R_{\mathrm{sdB}} = 0.30 \pm 0.03$ $R_{\odot}$. These values were used as priors in LCURVE (see section~\ref{Light curve Modeling}).


The light curve modeling gives a mass ratio $q = \frac{M_{\mathrm{sdB}}}{M_{\mathrm{WD}}} = 0.88 \pm 0.02$ with an sdB mass of $M_{\mathrm{sdB}} = 0.42 \pm 0.01$ $\mathrm{M_{\odot}}$ and a radius $R_{\mathrm{sdB}} =0.268\pm0.003$ $\mathrm{R_{\odot}}$ as well as a WD mass of $M_{\mathrm{WD}} = 0.48 \pm 0.01$ $\mathrm{M_\odot}$. The orbital inclination is $i = 48.1 \pm 0.2^\circ$ and the orbital separation is $a = 0.728 \pm 0.005$ $\mathrm{R_\odot}$. See Table~\ref{Results_table} for the full list of values obtained from the spectral analysis and light curve modeling. The corner plot summarizing the results of the MCMC and the resulting light curve model can be found in Appendix~\ref{C}.

\begin{table*}
\centering
\captionsetup{justification=centering}
\tiny
\begin{threeparttable}
\caption[]{Overview of ZTF J0007+4804 parameters}
\label{Results_table}
\renewcommand{\arraystretch}{1.4}

\begin{tabular}{lcccr}
    \hline
    \hline
    \noalign{\smallskip}
    Parameter               & Unit                          & Spectroscopic solution & Light curve solution   & Adopted value\\
    \hline
    Right ascension         & R.A                               & -                 & -                         & 00:07:42.63\\
    Declination             & Dec.                              & -                 & -                         & +48:04:14.45\\
    Parallax                &$\pi$ [mas]                        & -                 & -                         & 0.6695 $\pm$ 0.0312\\
    Distance                &$d$ [kpc]                          & -                 & -                         & 1.494 $\pm$ 0.069\\
    Apparent magnitude      &$m_\mathrm{V}$ [mag]                        & -                 & -                         & 15.037 $\pm$ 0.08\\
    Proper motion           &Pm RA [mas yr$^{-1}$]              & -                 & -                         & 2.31 $\pm$ 0.029\\
    Proper motion           &Pm Dec [mas yr$^{-1}$]             & -                 & -                         & -1.078 $\pm$ 0.023\\
    \hline
    \multicolumn{5}{c}{\text{Derived Parameters}} \\
    \hline
    Orbital period          &$P_{\mathrm{orb}}$ [minutes]                & -                 & 108.72 $\pm$ 0.01       & 108.72 $\pm$ 0.01\\
    Effective temperature   &$T_{\mathrm{eff}}$ [K]                      & 23\,500 $\pm$ 800   & fixed                     & 23\,500 $\pm$ 800\\
    System velocity         &$\gamma$ [km s$^{-1}$]             & 12.7 $\pm$ 0.6    & -                         & 12.7 $\pm$ 0.6\\
    Helium abundance        &log $y$ = log$\frac{n\mathrm{He}}{n\mathrm{H}}$      & -0.99 $\pm$ 0.07  & -                         & -0.99 $\pm$ 0.07\\
    Surface gravity         &$\log g$                            & 5.26 $\pm$ 0.11   & *5.185 $\pm$ 0.003         & 5.185 $\pm$ 0.003\\
    Rotational velocity     &$v_{\mathrm{rot}} \sin i$ [km s$^{-1}$]   & 132 $\pm$ 2       & *134 $\pm$ 2              & 134 $\pm$ 2\\
    RV semi-amplitude       &$K_{\mathrm{sdB}}$ [km s$^{-1}$]            & 194 $\pm$ 2       & *193 $\pm$ 2               & 193 $\pm$ 2\\
    Mass ratio              &$q = \frac{M_{\mathrm{WD}}}{M_{\mathrm{sdB}}}$       & -                 & 0.88 $\pm$ 0.02           & 0.88 $\pm$ 0.02\\
    sdB mass                &$M_{\mathrm{sdB}}$ [$M_\odot$]              & -                 & 0.42 $\pm$ 0.01           & 0.42 $\pm$ 0.01\\
    sdB radius              &$R_{\mathrm{sdB}}$ [$R_\odot$]              & 0.30 $\pm$ 0.03   & *0.268 $\pm$ 0.003         & 0.268 $\pm$ 0.003\\
    WD mass                 &$M_{\mathrm{WD}}$ [$M_\odot$]               & -                 & 0.48 $\pm$ 0.01           & 0.48 $\pm$ 0.01\\
    Orbital inclination     &$i$ [$^{\circ}$]                   & -                 & 48.1 $\pm$ 0.3            & 48.1 $\pm$ 0.3\\
    Separation              &$a$ [$R_\odot$]                    & -                 & 0.728 $\pm$ 0.005         & 0.728 $\pm$ 0.005\\
    \noalign{\smallskip}
    \hline
\end{tabular}

\begin{tablenotes}
\item[*] Values marked with an asterisk were computed using the values from the spectroscopic solution as priors.
\end{tablenotes}

\end{threeparttable}
\end{table*}

\begin{figure}
    \centering
    \includegraphics[width=\hsize]{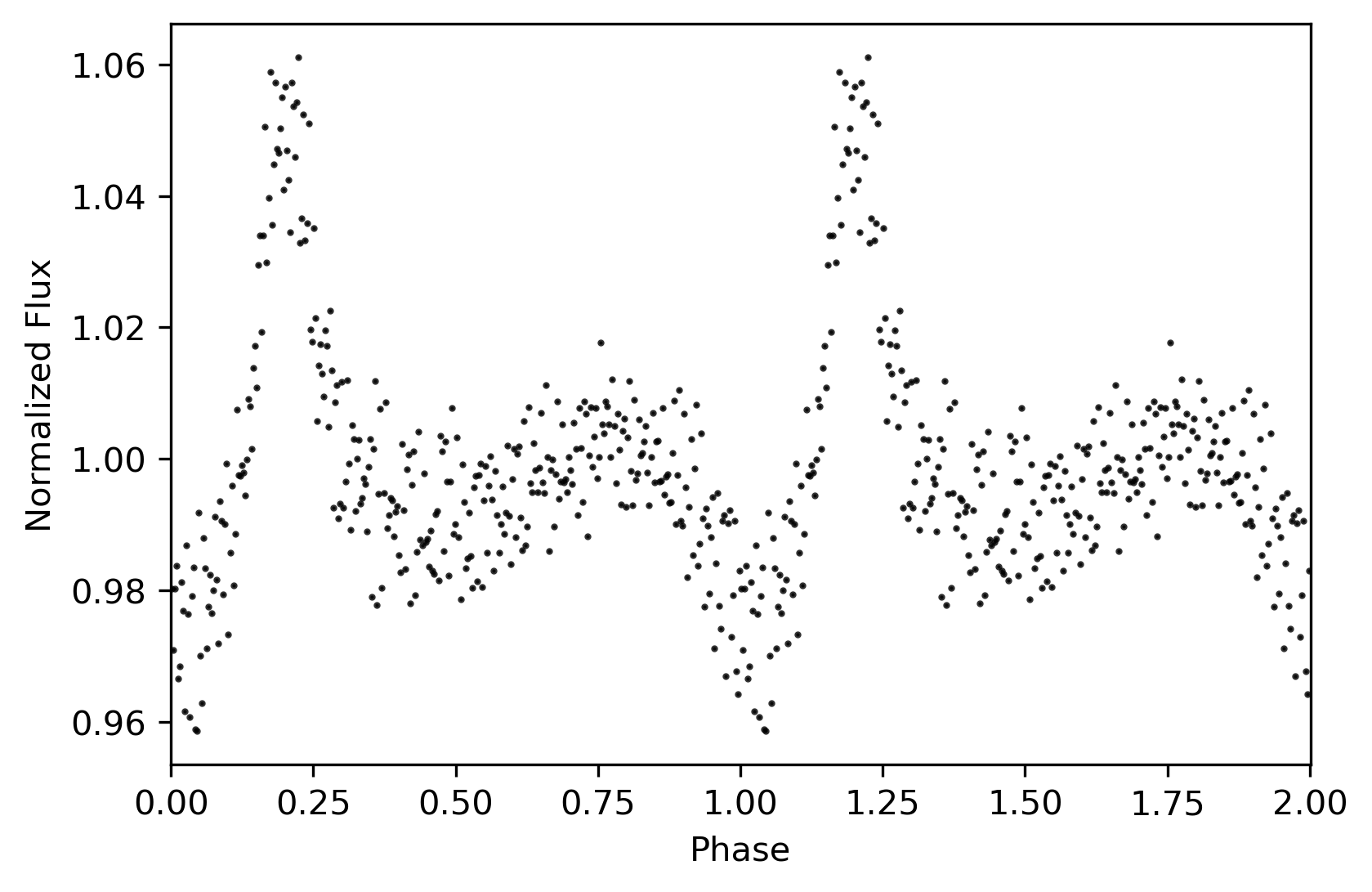}
    \caption{ZTF J0007+4804 TESS light curve folded on the outburst period. The data are binned over a single orbital period. The data are repeated for clarity.
      }
    \label{Outburst_Binned}
\end{figure}

\begin{figure*}
    \centering
    \includegraphics[width=\textwidth]{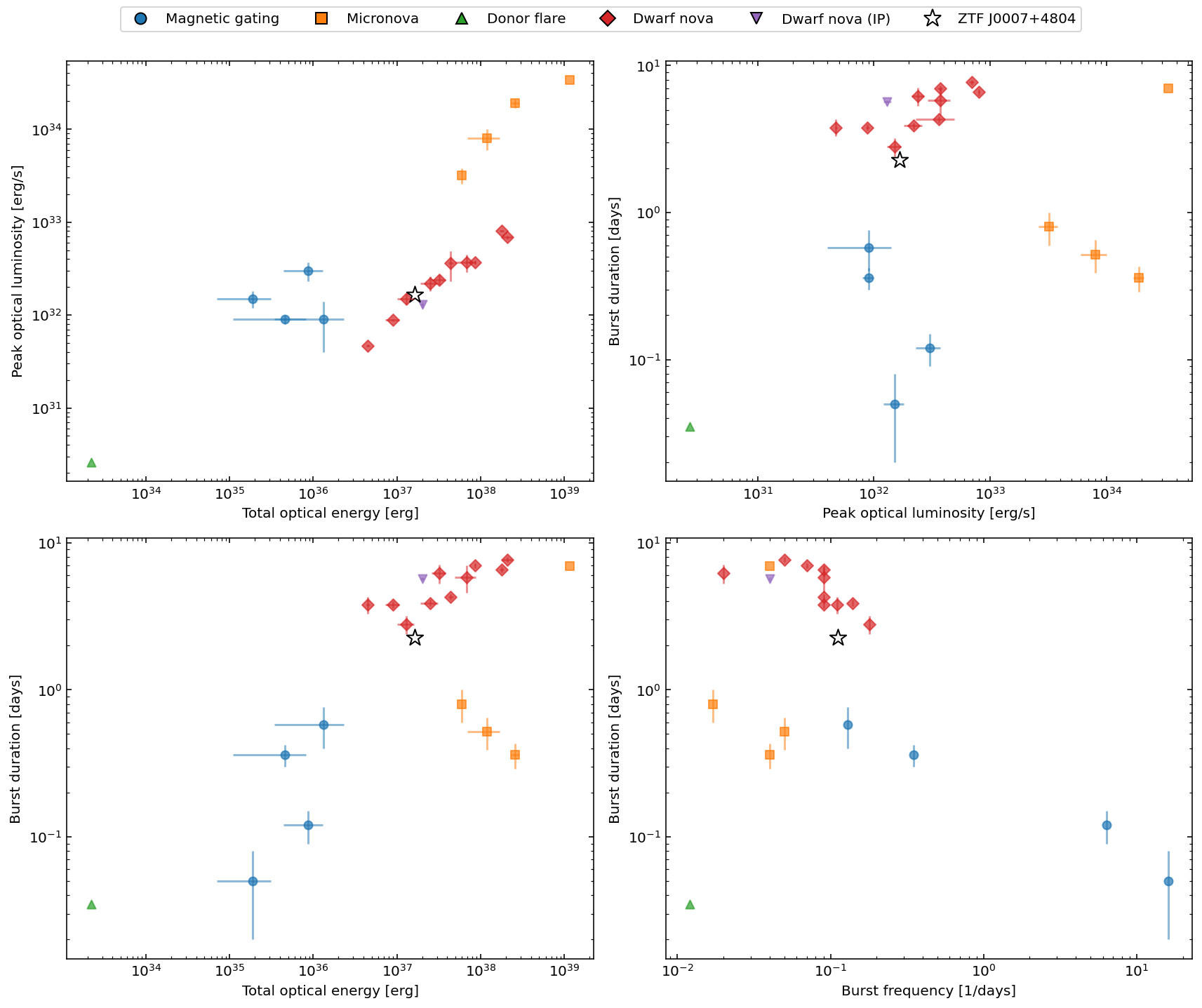}
    \caption{A recreation of Figure 2 from \cite{Ilkiwicz_2024} showing the properties of short bursts in CVs with their identification from the literature. ZTF J0007+4804 is included for comparison.}
    \label{dwarf_novae}
\end{figure*}

\begin{figure}
    \centering
    \includegraphics[width=\hsize]{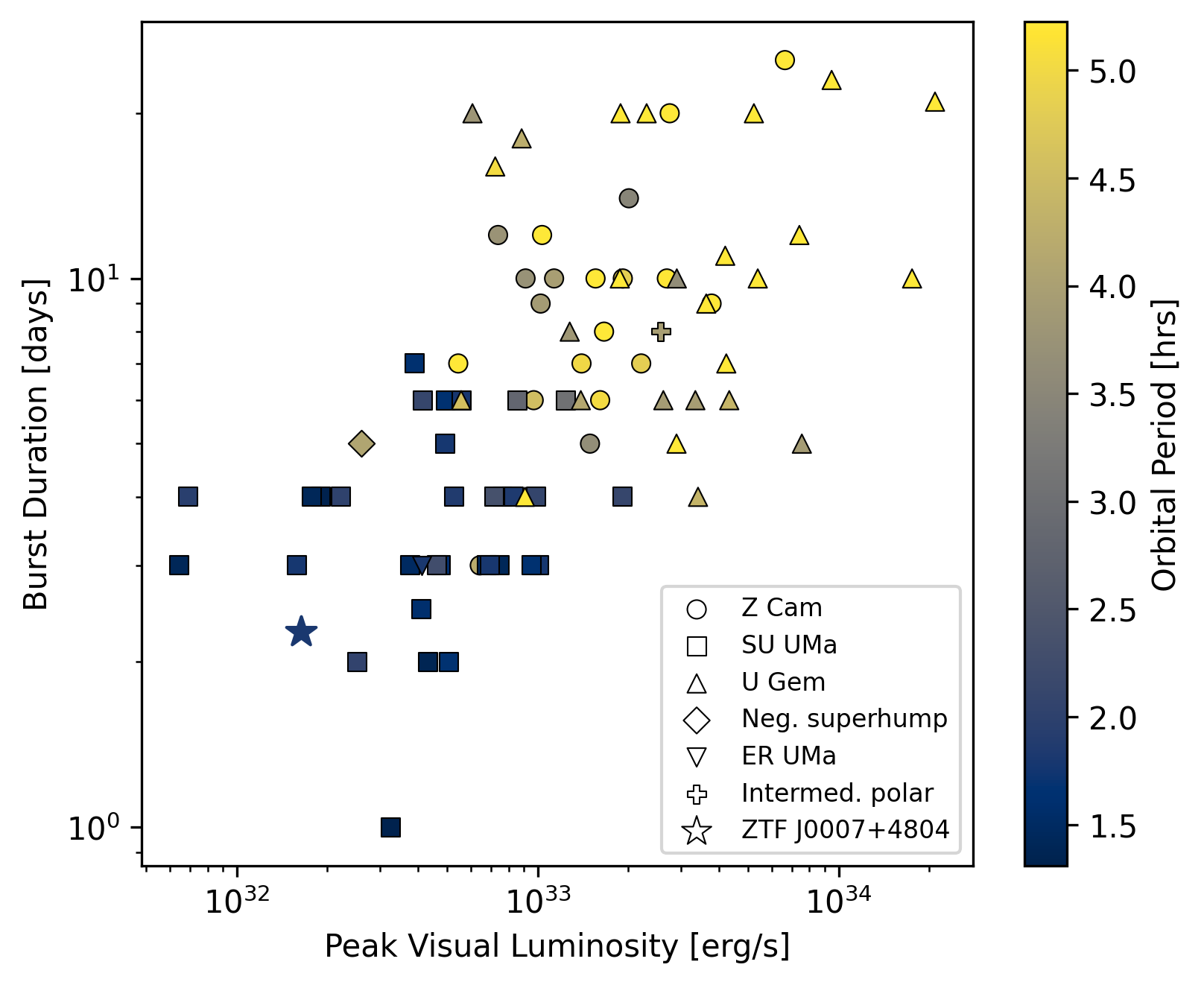}
    \caption{ZTF J0007+4804 outbursts compared to dwarf novae from \cite{Otulakowska-Hypka_2016}. Only normal outbursts are included in this plot. The superoutbursts from SU UMa systems are excluded.}
    \label{subtypes}
\end{figure}

\section{Discussion}\label{sec:discussion}

\subsection{Outbursts}
To classify the brightening events seen in ZTF J0007+4804, the diagnostic method described in \cite{Ilkiwicz_2024} was used. First, we can use the change in magnitude between the low and high states of the system to estimate the peak outburst luminosity in the ZTF $g$ band. $\Delta m_\mathrm{g}=m_{\mathrm{low}} -m_{\mathrm{high}}  \approx -0.05\ \textrm{mag}$. With an average low-state $g$-magnitude of $14.96 \ \textrm{mag}$ and extinction parameters E(g-r) = $0.10^{+0.2}_{-0.1}$ and $R_{\mathrm{g}} = 3.518$ \citep{Green_2019}, we get a corrected magnitude in the low state of $m_{\mathrm{low}} \approx 14.61 \ \textrm{mag}$. Using the zero-point spectral flux density for the ZTF $g$ band in the AB magnitude system of $f_{0,\lambda} = 4.75724 \times 10^{-9} \mathrm{erg \ s^{-1} \ cm^{-2}}$ and a wavelength range of 3676 -- 5613.82 $\AA$ we calculate the outburst luminosity in the $g$ band to be $L_{\mathrm{g}} \approx 1.7 \times 10^{32} \ \mathrm{erg/s}$. For a complete derivation, see Appendix \ref{B}. From the phase folded outburst light curve, the outburst duration (measured as the duration the flux is above the average quiescent value) was estimated to be $\Delta t_{\mathrm{out}} \approx 2.3 \ \mathrm{days}$ (Fig.~\ref{Outburst_Binned}). The total energy of the outburst was estimated simply as $E_{\mathrm{total}} \approx 0.5 \, \Delta t_{\mathrm{out}} L_{\mathrm{g}} \approx 1.6 \times 10^{37} \ {\mathrm{erg}}$.

Figure~\ref{dwarf_novae} shows the outburst properties of several types of brightening events considered in \cite{Ilkiwicz_2024}. Dwarf novae have a peak optical luminosity between $10^{31}\,\mathrm{erg/s}$ and $10^{33}\,\mathrm{erg/s}$ and a burst duration of 2-10 days. Their total outburst energy in optical bands is between $10^{36}\,\mathrm{erg}$ and $10^{38}\,\mathrm{erg}$, and their recurrence time is around 5 to 20 days. The properties of the brightening events of ZTF J0007+4804 match those of the dwarf novae described in \cite{Ilkiwicz_2024}. In addition, the expected mass transfer rate for compact sdB--WD binaries during the hydrogen accretion phase is $\dot{M} \sim 10^{-11}\ \mathrm{M_\odot}\ \textrm{yr}^{-1}$ to $10^{-10}\ \mathrm{M_\odot}\ \textrm{yr}^{-1}$ \citep{Bauer_2021} which is consistent with the average mass transfer rate for dwarf novae of about $10^{-12}$ $\mathrm{M_\odot}$ yr$^{-1}$ to $10^{-8}$ $\mathrm{M_\odot}$ yr$^{-1}$ \citep{Hameury_2020}. From this evidence we conclude that ZTF J0007+4804 exhibits dwarf nova outbursts due to the accretion of hydrogen rich material from the sdB onto the WD.

When we compare the outburst characteristics to a larger sample of known dwarf novae \citep{Otulakowska-Hypka_2016}, ZTF J0007+4804 aligns most closely with the SU UMa type dwarf novae regarding peak luminosity, burst duration, and orbital periods (Fig.~\ref{subtypes}). This subtype is characterized by the presence of superoutbursts in between regular outbursts which occur with a lower frequency than the normal outbursts, and last for one to two weeks. Superoutbursts in SU UMa dwarf novae are thought to be caused when the accretion disk that has undergone a normal thermal instability also expands to the 3:1 orbital resonance, becomes tidally unstable and eccentric, and then dissipates a large stored mass—producing a long, bright outburst with superhumps \citep{Osaki1989, Osaki1996, Osaki2005}. The 3:1 resonance requires $q = \frac{M_{\mathrm{donor}}}{M_{\mathrm{WD}}} \lesssim 0.25 - 0.33$ because only in sufficiently low–mass-ratio binaries is the accretion disk large enough to extend out to the 3:1 resonance radius before being tidally truncated by the companion \citep{WhitehurstKing1991}. This is possible in SU UMa systems as the donor stars are low mass main sequence stars with $M_2\approx0.08\text{–}0.20\,\mathrm{M_\odot}$ \citep{knigge2006, knigge2011}. Despite its similarities to SU UMa type dwarf novae, we do not find evidence for a superoutburst in the TESS lightcurve (Figure~\ref{Tess_Lightcurve}) as well as ZTF, Gaia DR3, or ASAS-SN \citep{Shappee_2014,Kochanek_2017} for ZTF\,J0007+4804. However, this is also not expected as the donor star in ZTF\,J0007+4804 is an sdB and the mass ratio of the system is significantly larger compared to typical SU UMa dwarf novae.

\begin{figure}
   \centering
   \includegraphics[trim={0.5cm 0cm 1.0cm 0cm}, clip, width=\hsize]{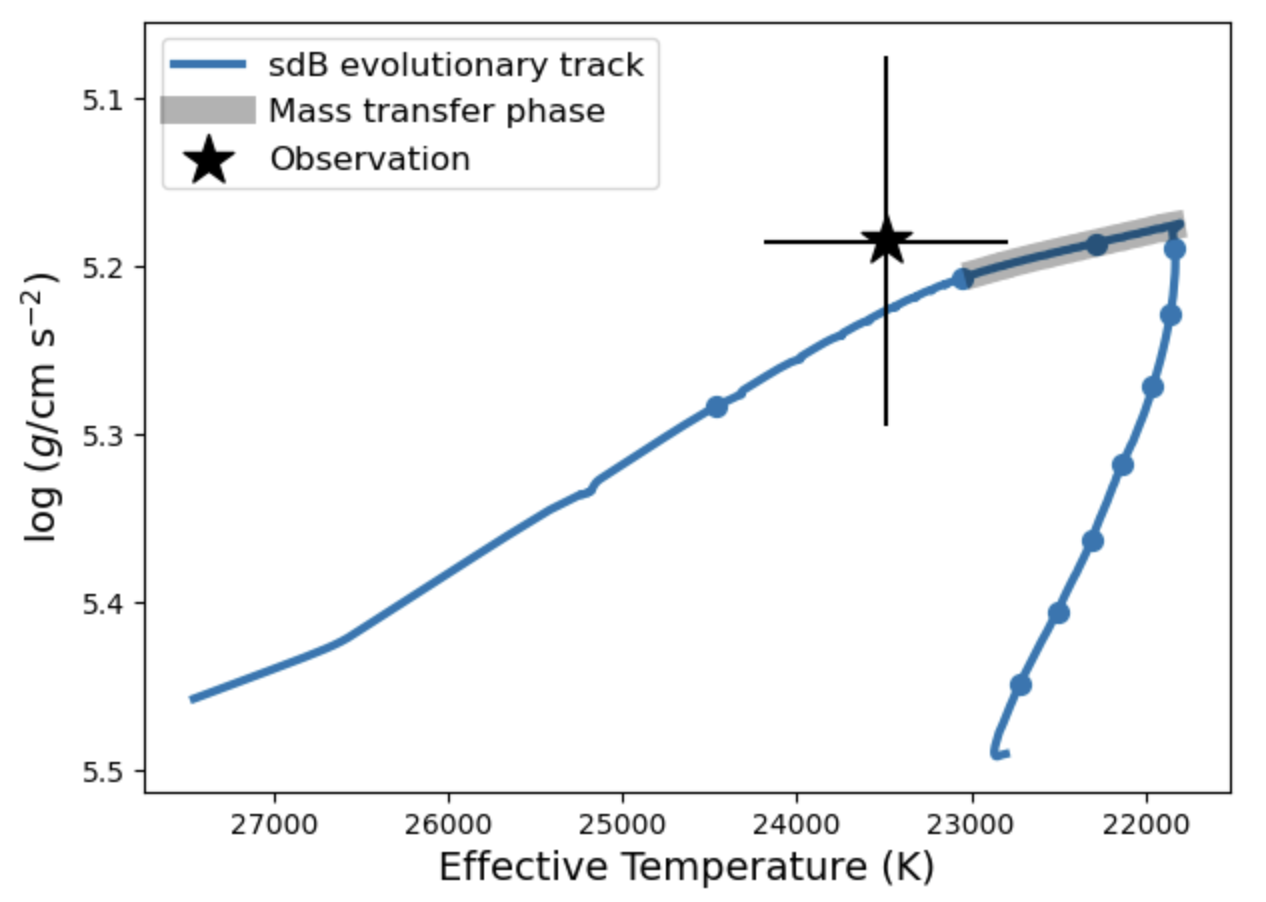}
    \caption{Evolutionary track of a $0.42\,M_\odot$ sdB with a $3\times10^{-3}\,M_\odot$ hydrogen envelope shown in the $T_{\rm eff}-{\rm log}\,g$ parameter space. The track (blue) begins in the bottom right, evolving toward lower ${\rm log}\,g$ until filling its Roche lobe and initiating mass transfer (grey). Blue points on the track mark 20 Myr timesteps, while the black star shows the observed values with errors. The mass transfer region represents $\dot{M} > 10^{-12}\,M_\odot\,{\rm yr}^{-1}$.}
    \label{evol_track}
\end{figure}

\begin{figure}
    \centering
    \includegraphics[width=\hsize]{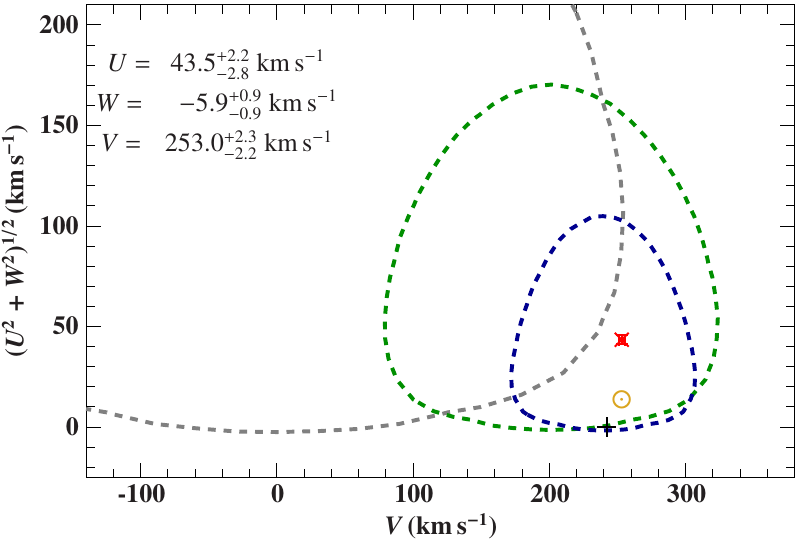}
    \caption{Toomre diagram of ZTF J0007+4804. $U$ is the component of the Galactic orbit toward the Galactic center, $V$ is in the direction of Galactic rotation, and $W$ is parallel to the north Galactic pole. 2$\sigma$ contours are plotted for the thin disk (blue), thick disk (green), and halo (grey), based on data from \cite{Robin_2003}. The solar orbit is marked as a yellow circle. 
              }
    \label{Toomre}
\end{figure}
   

\subsection{Evolutionary history}


ZTF\,J0007+4804 has been measured to host an sdB star with a mass of
$M_{\mathrm{sdB}} = 0.42 \pm 0.01\,\mathrm{M_{\odot}}$ and a white dwarf companion with
$M_{\mathrm{WD}} = 0.48 \pm 0.01\,\mathrm{M_{\odot}}$. To understand the evolutionary history of this system, two key aspects must be considered. First, the sdB has a non-canonical mass and can only be explained if it has evolved from a main-sequence progenitor of approximately $2.0$–$2.1\,\mathrm{M_{\odot}}$ or $3.1$–$3.2\,\mathrm{M_{\odot}}$ \citep{Bauer_2021}. Second, the WD companion is significantly less massive than expected for progenitors with masses $>1\,\mathrm{M_{\odot}}$ and is also lighter than the WDs found in other compact sdB+WD systems such as PTF\,J2238 \citep{Kupfer_2022} or CD$-30^\circ11223$ \citep{Geier_2013}. We therefore propose the following evolutionary scenario for ZTF\,J0007+4804, which explains all observed properties and closely resembles the channel discussed by \citet{hernandez2026} with a small modification to match the low WD companion mass.

The system likely began as a binary consisting of a $\approx2\,\mathrm{M_{\odot}}$ main-sequence star (the future sdB; see Sect.\,\ref{sec:future}) and a slightly less massive companion in an orbit of a few days. The sdB progenitor evolved first and initiated stable mass transfer onto the companion. By the end of this phase, the sdB had formed with its observed mass of $0.42\,\mathrm{M_{\odot}}$, and the orbit had widened substantially, consistent with the first stable Roche-lobe overflow channel described by \citet{Han_2002, Han_2003, hernandez2026}. The companion accreted material during this phase and became a $\approx2.5$–$3\,\mathrm{M_{\odot}}$ star. It then evolved off the main sequence and filled its Roche lobe while the sdB was still burning helium in its core. Owing to the large mass ratio at this stage, the mass transfer became dynamically unstable and triggered a common-envelope phase.

To produce the observed low-mass WD companion, this common-envelope event must have occurred early in the AGB evolution of the companion. Figure 23 of \citet{choi2016} shows the growth of the core mass during the thermal-pulse AGB phase as a function of initial mass. For a star with an initial mass of $\approx3\,\mathrm{M_{\odot}}$, the core mass is about $\approx0.45\,\mathrm{M_{\odot}}$ at the first thermal pulse and increases to $\approx0.65\,\mathrm{M_{\odot}}$ during the AGB phase. Therefore, if the common envelope is initiated early in the thermal-pulse phase, the remnant will be a low-mass WD, consistent with what is observed in ZTF\,J0007+4804.

\subsection{Current phase and future evolution}\label{sec:future}

To test the feasibility of a mass-transferring sdB consistent with the observed $T_{\rm eff}-{\rm log}\,g$, we attempted simulating ZTF\,J0007+4804 using the 1-D stellar evolution code MESA, version 23.05.1\citep{2011MESA,2013MESA,2015MESA,2018MESA,2019MESA,2023MESA}. For modeling the sdB, we followed the same approach as \citet{2024Deshmukh}, most notably including predictive mixing and element diffusion effects (see Sect.\,3 in their paper for more details). Given the mass constraints from Table\,\ref{Results_table}, we explored sdBs with progenitor masses between $2.0-2.1\,\mathrm{M_\odot}$ and envelope masses between $10^{-4}-10^{-2}\,\mathrm{M_\odot}$. We further modeled the sdB-WD binary using the binary module in MESA. The WD was assumed to be a point mass with $M_2 = 0.48\,\mathrm{M_\odot}$, while the mass transfer was set to be fully conservative.

We found a configuration compatible within error bars of the observed properties of the system, where (i) the sdB has a mass $M_1=0.42\,\mathrm{M_\odot}$ with a $3\times10^{-3}\,\mathrm{M_\odot}$ envelope, formed from a $2.05\,\mathrm{M_\odot}$ progenitor; (ii) the mass transfer initiates during core-helium burning of the sdB; (iii) the evolutionary track passes through the observed $T_{\rm eff}-{\rm log}\,g$ error region during mass transfer, when the orbital period is around 115 minutes; (iv) the corresponding mass transfer rate is $\dot{M} \approx 1.6\times10^{-11} \mathrm{M_\odot}\,{\rm yr}^{-1}$, with mostly hydrogen-rich material from the sdB envelope being transferred to the WD. The sdB evolutionary track is shown in Fig.\,\ref{evol_track}. Since a $0.42\,\rm{M_\odot}$ sdB can also descend from a $3.1-3.2\,\rm{M_\odot}$ \citep{Bauer_2021}, we explored the scenario but found the sdBs to be too compact to satisfy the ${\rm log}\,g$ constraint.

As the sdB exhausts helium in its core, the system is likely to evolve into a double C/O WD binary with comparable masses. Based on the current orbital period and masses, they are expected to merge in approximately 226 Myr due to gravitational wave emission, although later mass transfer episodes can modify this time slightly. The merger is expected to produce a hot, differentially rotating merger remnant consisting of a central C/O core surrounded by a helium-rich envelope and accretion disk after the lower-mass WD is tidally disrupted; this object then thermally relaxes over $\approx10^5$ yr into a single massive hydrogen-deficient white dwarf, potentially appearing as a DO/DB or hot-DQ white dwarf \citep{schwab2021, schwab2021a}. However a thermonuclear explosion cannot be ruled out \citep{shen2024}.



\subsection{Kinematics}

If the sdB star has evolved from a $2.0-2.1\,\rm{M_\odot}$ star, the system has to be part of a young population. To put constraints on the population origin of ZTF\,J0007+4804, we calculated its kinematics. The Galactic orbit of ZTF\,J0007+4804 was calculated using the system velocity $\gamma$, and the parallax and proper motions from Gaia DR3 to calculate the Galactic velocity components $U$, $V$, and $W$. For more details see \cite{Irrgang_2014_phd}. Model I from \cite{Irrgang_2013} was used for the Milky Way mass model, and the orbit was traced backwards for 5 Gyrs. 

The Toomre diagram (Figure~\ref{Toomre}) places ZTF\,J0007+4804 in the overlap region of the thin disk and thick disk. The z-component of the Galactic orbit stays below $0.6$ kpc, suggesting that the system is a member of the thin disk \citep{Richter_2007}.


\subsection{Comparison to similar systems}
ZTF J0007+4804 is consistent with a Roche lobe filling compact sdB--WD binary. Unlike the three previously discovered Roche lobe filling systems, ZTF J0007+4804 does not show disk eclipses. However, it is difficult to explain the observed outbursts without the presence of a disk, especially since the diagnostic method of \cite{Ilkiwicz_2024} places the system with other dwarf novae. Furthermore, the light curve model is consistent with a Roche Lobe filling model and gives an inclination for the system of $i \approx 48^\circ$ which excludes the possibility of disk eclipses. Like the three previously known systems, ZTF J0007+4804 does not show spectral lines from the accretion disk. This can be explained by the low flux ratio of the quiescent disk to the sdB, though emission lines may be visible at the beginning of an outburst, before the disk becomes optically thick.

Despite having a lower temperature than the other known Roche lobe filling systems, ZTF J0007+4804 ($M_g=3.76\ \mathrm{mag}$, $T_{\rm{eff}} = 23\,500\ \mathrm{K}$, $R=0.268\ \mathrm{R_\odot}$) is brighter than both ZTF J213056.71+442046.5 ($M_g = 4.3\ \mathrm{mag}$, $T_{\rm{eff}}=42\,400\ \mathrm{K}$, $R=0.125\ \mathrm{R_\odot}$) \citep{Kupfer_2020b} and ZTF J205515.98+465106.5 ($M_g=4.4\ \mathrm{mag}$, $T_{\rm{eff}}=33\,700\ \mathrm{K}$, $R=0.17\ \mathrm{R_\odot}$) \citep{Kupfer_2020a} in visual bands. This is due to both ZTF J0007+4804 having a larger radius and that most of the excess energy from the higher temperatures is in the ultraviolet range. This suggests that dwarf nova outbursts similar to those seen in ZTF J0007+4804 should be visible in ZTF J213056.71+442046.5 and ZTF J205515.98+465106.5 if they exist.

Even though the majority of hot subdwarf stars are believed to have formed through low mass progenitors, all known Roche lobe filling hot subdwarf-WD binaries have been modeled to have come from stripped stars with progenitor masses $\gtrsim2$ $\mathrm{M}_\odot$. This is in line with theory by \cite{Han_2003} that predicts that the most compact hot subdwarf-WD systems should come from higher mass systems. ZTF J0007+4804 is the first Roche lobe filling system that is in the helium core burning phase of its evolution. ZTF J213056.71+442046.5 and J205515.98+465106.5 are in a short hydrogen shell burning phase near the end of the hot subdwarf's lifetime that lasts a factor of $\approx10-100$ times shorter than the helium core burning phase. SMSS J192054.50--200135.5 is suspected of being a stripped AGB star also undergoing shell burning, though it is expected to have been born as a shell burning object. Statistically, we would expect to see many more systems with helium core burning hot subdwarfs than shell burning hot subdwarfs. This discrepancy could be due to a selection bias. Hot subdwarfs in the hydrogen shell burning phase are typically sdO stars, and they are more easily separated from the main sequence when making color selection cuts of large surveys when searching for short period sdB-WD binaries. Cooler systems like ZTF J0007+4804 may be missed in these color selection cuts (being hidden in the main sequence) if they are more heavily reddened like ZTF J205515.98+465106.5. The kinematics of ZTF J0007+4804, ZTF J213056.71+442046.5 and ZTF J205515.98+465106.5 place these systems in the thin disk. This, combined with the high progenitor masses from MESA modeling, indicates that all of these systems come from a young stellar population. SMSS J192054.50–200135.5 is an outlier in this case, having been modeled from a high mass progenitor, but with kinematics pointing to an origin in the Galactic halo. This makes the origin of SMSS J192054.50–200135.5 unclear.

\section{Summary and Conclusions}
\label{Conclusion}
In this work, we performed a detailed analysis of the compact binary system ZTF J0007+4804. Using data from ZTF and TESS we find the orbital period to be $P_\mathrm{orb} = 108.72 \pm 0.01$ minutes. Analysis of time-resolved spectroscopy was combined with light curve modeling from LCURVE, confirming the system as an sdB--WD binary undergoing Roche lobe overflow, the fourth such system discovered to date. The primary is an sdB with an effective temperature of  $T_{\rm{eff}}=23\,500 \pm 800\,\rm{K}$, surface gravity of $\log g=5.185 \pm 0.003$, and a mass of $M_{\rm{sdB}}=0.42\pm0.01\,\rm{M_\odot}$. The white dwarf has a mass of $M_{\rm{WD}}=0.48\pm0.01\,\rm{M_{\odot}}$. ZTF J0007+4804 is also the first known sdB--WD system that exhibits periodic outbursts. Using TESS data, we performed a Lomb-Scargle analysis to determine the outburst period to be $P_\mathrm{out} \approx 9$ days. We characterized the outbursts using the method described in \cite{Ilkiwicz_2024}, and found the system to be consistent with SU UMa type dwarf novae. X-ray observations were performed with Swift, setting an upper limit on the X-ray luminosity at $3\times10^{31}$erg/sec which is consistent with a non-detection for typical dwarf novae. Using MESA modeling, we find the sdB to be in the helium core burning phase in contrast to the other known Roche lobe filling hot subdwarf--WD binaries. From the same MESA modeling we estimate the mass transfer rate of the system to be $\dot{M} \approx 1.6\times10^{-11} \rm{M_\odot}\,{\rm yr}^{-1}$. After approximately 226 Myr the system will merge due to gravitational wave emission and likely result in a single massive hydrogen-deficient white dwarf, but a thermonuclear explosion cannot be ruled out. 

\begin{acknowledgements}
      This research was supported by Deutsche Forschungsgemeinschaft  (DFG, German Research Foundation) under Germany’s Excellence Strategy - EXC 2121 "Quantum Universe" – 390833306. Co-funded by the European Union (ERC, CompactBINARIES, 101078773). Views and opinions expressed are however those of the author(s) only and do not necessarily reflect those of the European Union or the European Research Council. Neither the European Union nor the granting authority can be held responsible for them.
      For this work the HPC-cluster Hummel-2 at University of Hamburg was used. The cluster was funded by Deutsche Forschungsgemeinschaft (DFG, German Research Foundation) – 498394658. VSD and ULTRASPEC are supported by STFC grant  ST/Z000033/1. AP acknowledges support from the FWO under grant agreement No. 11M8325N (PhD Fellowship). 
      
      Based on observations obtained with the Samuel Oschin 48- inch Telescope at the Palomar Observatory as part of the Zwicky Transient Facility project. ZTF is supported by the National Science Foundation under grant No. AST-1440341 and a collaboration including Caltech, IPAC, the Weizmann Institute for Science, the Oskar Klein Center at Stockholm University, the University of Maryland, the University of Washington, Deutsches Elektronen-Synchrotron and Humboldt University, Los Alamos National Laboratories, the TANGO Consortium of Taiwan, the University of Wisconsin at Milwaukee, and Lawrence Berkeley National Laboratories. Operations are conducted by COO, IPAC, and UW. 
      
      This research has made use of the Keck Observatory Archive (KOA), which is operated by the W. M. Keck Observatory and the NASA Exoplanet Science Institute (NExScI), under contract with the National Aeronautics and Space Administration. Some of the data presented herein were obtained at the W. M. Keck Observatory, which is operated as a scientific partnership among the California Institute of Technology, the University of California, and the National Aeronautics and Space Administration. The Observatory was made possible by the generous financial support of the W. M. Keck Foundation. The authors wish to recognize and acknowledge the very significant cultural role and reverence that the summit of Maunakea has always had within the indigenous Hawaiian community. We are most fortunate to have the opportunity to conduct observations from this mountain. 
      
      This paper includes data collected with the TESS mission, obtained from the MAST data archive at the Space Telescope Science Institute (STScI). Funding for the TESS mission is provided by the NASA Explorer Program. STScI is operated by the Association of Universities for Research in Astronomy, Inc., under NASA contract NAS 5–26555. 
      
      We acknowledge the use of public data from the Swift data archive.
\end{acknowledgements}

%
%


\bibliographystyle{aa} 
\bibliography{bibliography} 

\clearpage
\appendix
\section{Quiescent Flux Ratio}
\label{A}
We can estimate an upper limit for the flux ratio of the quiescent disk to the sdB $\frac{F_{\mathrm{dq}}}{F_{\mathrm{sdB}}}$ by using the flux ratio of the quiescent disk to the outbursting disk $\frac{F_{\mathrm{do}}}{F_{\mathrm{dq}}}$. Subscripts $d$ and $s$ denote the flux of the disk or system, and subscripts $o$ and $q$ denote the flux in outburst or quiescence. We start with finding the flux ratio of the quiescent and outburst states of the disk.
\begin{equation}
\Delta m_\mathrm{d} = -2.5\log\left(\frac{F_{\mathrm{do}}}{F_{\mathrm{dq}}}\right)
\label{A.1}
\end{equation}
\begin{equation}
10^\frac{-\Delta m_\mathrm{d}}{2.5} = \frac{F_{\mathrm{do}}}{F_{\mathrm{dq}}} \equiv \mathcal{R}_{\mathrm{d}}
\label{A.2}
\end{equation}
$\Delta m_\mathrm{d}$ for dwarf novae ranges between -1.8 and -5 during normal outbursts, leading to $5 < \mathcal{R}_{\mathrm{d}} < 600$. 

Similarly, we can write the change in magnitude of the system as a whole as:
\begin{equation}
\Delta m_\mathrm{s} = -2.5 \log\left(\frac{F_{\mathrm{so}}}{F_{\mathrm{sq}}}\right)
\label{A.3}
\end{equation}
The magnitude of the entire system in outburst will be from the flux of the sdB $F_{\mathrm{sdB}}$ and from the flux of the disk in outburst $F_{\mathrm{do}}$. Conversely, when the system is in quiescence, the magnitude of the system will be from the quiescent flux of the disk $F_{\mathrm{dq}}$ while the sdB flux $F_{\mathrm{sdB}}$ will remain the same.
\begin{equation}
10^{\frac{-\Delta m_\mathrm{s}}{2.5}} = \frac{F_{\mathrm{so}}}{F_{\mathrm{sq}}} \equiv \mathcal{R}_{\mathrm{s}} = \frac{F_{\mathrm{sdB}} + F_{\mathrm{do}}}{F_{\mathrm{sdB}} + F_{\mathrm{dq}}}
\label{A.4}
\end{equation}
From equation~\ref{A.2} we can substitute $\mathcal{R}_{\mathrm{d}}F_{\mathrm{dq}}$ for $F_{\mathrm{do}}$
\begin{equation}
\mathcal{R}_{\mathrm{s}} = \frac{F_{\mathrm{sdB}} + \mathcal{R}_{\mathrm{d}}F_{\mathrm{dq}}}{F_{\mathrm{sdB}} + F_{\mathrm{dq}}}
\label{A.5}
\end{equation}
Now solve for $\frac{F_{\mathrm{dq}}}{F_{\mathrm{sdB}}}$
\begin{equation}
\frac{F_{\mathrm{dq}}}{F_{\mathrm{sdB}}} =\frac{(\mathcal{R}_{\mathrm{s}} - 1)}{(\mathcal{R}_{\mathrm{d}} -\mathcal{R}_{\mathrm{s}})}
\label{A.9}
\end{equation}
The flux of ZTF J0007+4804 increases by about 10\% during outburst, meaning $\mathcal{R}_{\mathrm{s}} \approx1.1$. Combining with $\mathcal{R}_\mathrm{d} \approx 5$ this leads to $\frac{F_{\mathrm{dq}}}{F_{\mathrm{sdB}}} \lesssim 0.03$

\section{Outburst Optical Luminosity}
\label{B}
The change in magnitude between the low and high states of the system is defined by
\begin{equation}
\Delta m = m_{\mathrm{high}} - m_{\mathrm{low}} = -2.5 \log \left(\frac{F_{\mathrm{high}}}{F_{\mathrm{low}}} \right)
\label{B.1}
\end{equation}
Where the flux in the high state is the sum of the flux in the low state and the flux from the outburst
\begin{equation}
F_{\mathrm{high}} = F_{\mathrm{low}} + F_{\mathrm{out}}
\label{B.2}
\end{equation}
Plugging into \ref{B.1} and solving for $F_{\mathrm{out}}$ we get 

\begin{equation}
F_{\mathrm{out}} = F_{\mathrm{low}}\left(10^{-\frac{\Delta m}{2.5}}-1 \right)
\label{B.3}
\end{equation}
To find $F_{\mathrm{low}}$ we start with the equation for the magnitude 
\begin{equation}
m_{\mathrm{low}} = -2.5 \log \left(\frac{f_{\mathrm{low}}}{f_{0, \lambda}} \right)
\label{B.4}
\end{equation}
\begin{equation}
f_{\mathrm{low}} = f_{0,\lambda} \cdot 10^{-\frac{m_{\mathrm{low}}}{2.5}}
\label{B.5}
\end{equation}
where $f_{0,\lambda}$ is the filter's zero-point spectral flux density with units erg s$^{-1}$ cm$^{-2}$ $\AA ^{-1}$. In the AB magnitude system, $f_{0,\nu}$ is defined to be 3631 Jy. We can convert to $f_{0,\lambda}$ by
\begin{equation}
f_{0,\lambda} =2.9979246 \times10^{-5}\cdot \lambda_{\mathrm{pivot}}^{-2} * f_{0,\nu}
\label{B.6}
\end{equation}
where $\lambda_{\mathrm{pivot}}$ is defined from the filter's transmission function $T(\lambda)$
\begin{equation}
\lambda_{\mathrm{pivot}} = \sqrt{\frac{\int{T(\lambda) \,  \lambda\ d\lambda}}{\int{T(\lambda) \, \lambda^{-1} \ d\lambda}}}
\label{B.7}
\end{equation}
The spectral flux density $f$ in a photometric band as derived from the magnitude is the average value of the spectrum over the band width $\Delta \lambda$. Thus, by the mean value theorem, the total flux of the low state in a specific band is
\begin{equation}
F_{\mathrm{low}} = \int_{\lambda_{1}}^{\lambda_2} f_{\mathrm{low}} \ d\lambda = f_{\mathrm{low}} \cdot \Delta \lambda
\label{B.8}
\end{equation}
Plugging in $f_{\mathrm{low}}$ from equation \ref{B.5} we get
\begin{equation}
F_{\mathrm{low}} = 10^{-\frac{m_{\mathrm{low}}}{2.5}} \cdot f_{0,\lambda} \cdot \Delta \lambda
\label{B.9}
\end{equation}
Now plug \ref{B.9} into \ref{B.3} to get
\begin{equation}
F_{\mathrm{out}} = \left(10^{-\frac{\Delta m}{2.5}} - 1 \right)\cdot 10^{-\frac{m_{\mathrm{low}}}{2.5}} \cdot f_{0,\lambda}\cdot \Delta \lambda
\label{B.10}
\end{equation}
Assuming an isotropic source, we can estimate the luminosity of the outburst in a specific band to be
\begin{equation}
L_{\mathrm{out}} = 4\pi d^2F_{\mathrm{out}}
\label{B.11}
\end{equation}
\begin{equation}
L_{\mathrm{out}} = 4\pi d^2\left(10^{-\frac{\Delta m}{2.5}} - 1 \right)10^{-\frac{m_{\mathrm{low}}}{2.5}}\cdot f_{0,\lambda}\cdot \Delta \lambda
\label{B.12}
\end{equation}

\section{Additional Plots}
\label{C}

\begin{figure*}
   \centering
   \includegraphics[width=\hsize]{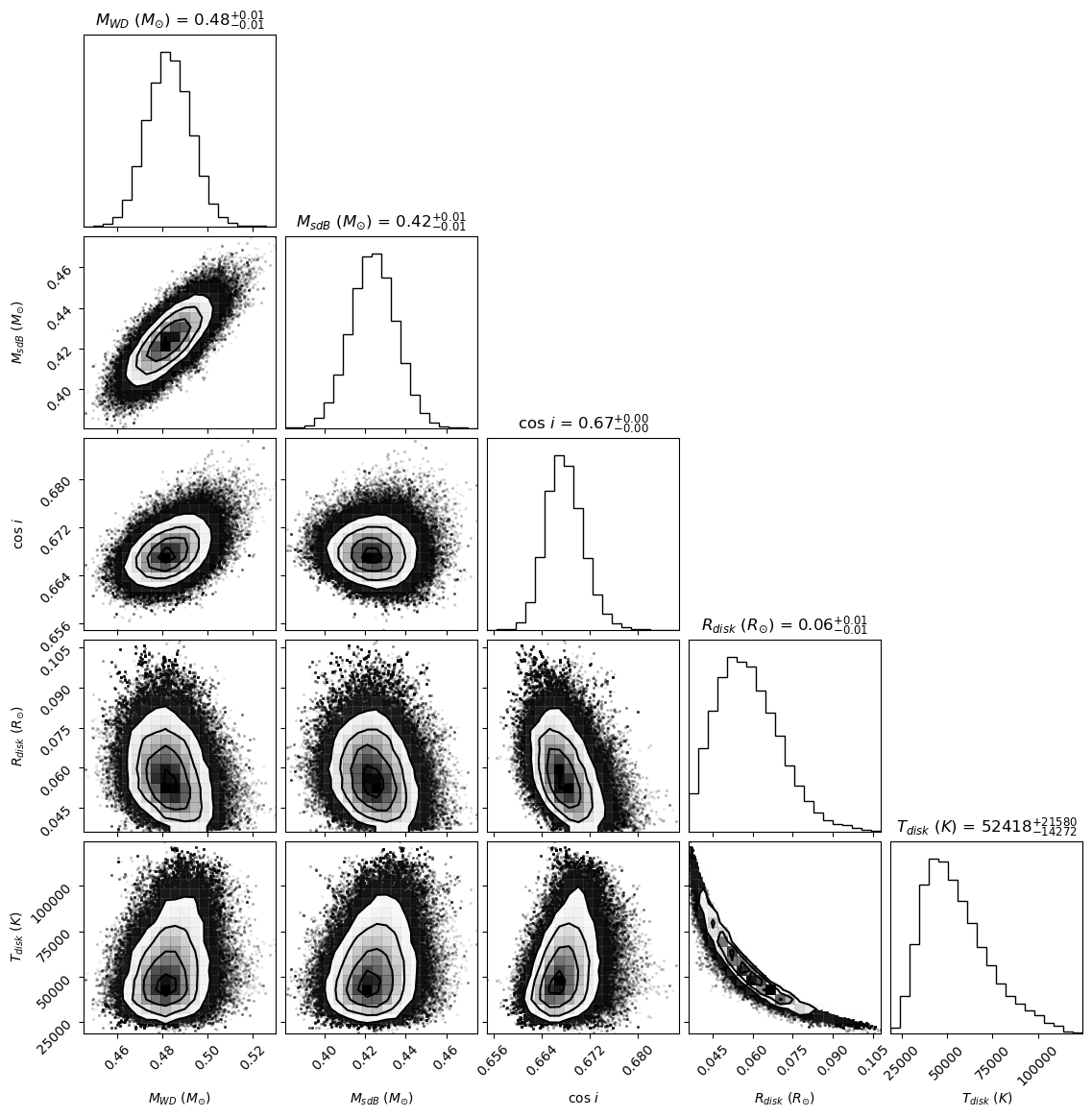}
   \caption{Corner plot of the parameters directly fit by the MCMC for ZTF J0007+4804.}
   \label{Corner}
\end{figure*}


\begin{figure*}
    \centering
    \includegraphics[width=\textwidth]{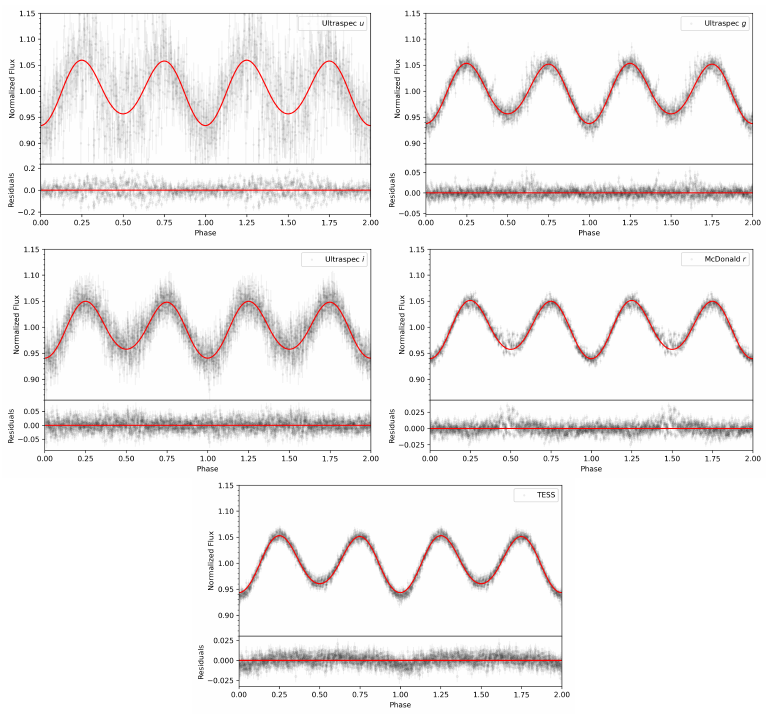}
    \caption{Top panels: Light curve model of ZTF J0007+4804 (red) compared to each photometric data set (gray). Bottom panels: The residuals of the fits.}
    \label{lcurve_fits}
\end{figure*}

\end{document}